\documentclass[notitlepage,a4paper,aps,prd,onecolumn,superscriptaddress,nofootinbib,groupedaddress,longbibliography]{revtex4-1}
\usepackage{amsmath}
\usepackage{amsfonts}
\usepackage{amssymb}
\usepackage[utf8]{inputenc}
\usepackage[colorlinks=true]{hyperref}
\usepackage{tikz}

\newcommand{\lc}[1]{\overset{\circ}{#1}\vphantom{#1}}

\begin{document}

\title{Polarization of gravitational waves in general teleparallel theories of gravity}
\author{Manuel Hohmann}
\email{manuel.hohmann@ut.ee}
\affiliation{Laboratory of Theoretical Physics, Institute of Physics, University of Tartu, W. Ostwaldi 1, 50411 Tartu, Estonia}

\begin{abstract}
We determine the possible gravitational wave  polarizations in two general classes of teleparallel gravity theories, using the metric and symmetric teleparallel geometries. For this purpose we apply the Newman-Penrose formalism, and find that depending on the choice of parameters, the \(\mathrm{E}(2)\) class of the theories is one of \(\mathrm{N}_2, \mathrm{N}_3, \mathrm{III}_5, \mathrm{II}_6\), corresponding to two to six polarizations, where all of them include the two tensor polarizations known from general relativity. We also find classes of theories apart from general relativity which yield the same polarizations.
\end{abstract}

\maketitle

\section{Introduction}\label{sec:intro}
With the first joint observation of a binary black hole merger at LIGO and VIRGO it has become possible for the first time to measure the polarization of gravitational waves~\cite{Abbott:2017oio}. While there are strong hints towards the existence of two tensor polarizations, the existence of further modes cannot yet be fully excluded. With the advent of more gravitational wave detectors, and the increasing statistics of such observations, one may expect more stringent bounds on their existence, and thus gain another useful tool for testing gravity theories.

The idea to use gravitational wave polarizations as a method for testing gravity theories has been developed already when the first attempts to observe gravitational waves have been taken. A very useful tool to describe the effect of a gravitational wave is the Newman-Penrose formalism, which employs a double null basis of the tangent space to describe the components of the Riemann tensor~\cite{Newman:1961qr}. This formalism is at the heart of the \(\mathrm{E}(2)\) classification scheme, which classifies gravity theories by determining which polarizations are compatible with the field equations of the theory~\cite{Eardley:1973br,Eardley:1974nw}.

While the curvature of the Levi-Civita connection is carrying the gravitational interaction in the standard formulation of general relativity and many of its extensions, there exist equivalent formulations of general relativity, in which gravity is described by torsion~\cite{Moller1961,Cho:1975dh,Maluf:2013gaa} or non-metricity~\cite{Nester:1998mp,Adak:2005cd}, which allow for different extensions and modifications. The aim of this article is to apply the Newman-Penrose formalism to two general classes of theories based on these geometries, and to determine their \(\mathrm{E}(2)\) classes.

We start with a brief review of the Newman-Penrose formalism and the \(\mathrm{E}(2)\) classification of gravitational wave polarizations in section~\ref{sec:newpen}. We then apply this formalism to metric teleparallel gravity in section~\ref{sec:torsion} and to symmetric teleparallel gravity in section~\ref{sec:nonmet}, in order to derive their \(\mathrm{E}(2)\) classes. We end with a conclusion in section~\ref{sec:conclusion}. Throughout this article Latin indices are Lorentz indices, while Greek indices are spacetime indices. Indices are raised and lowered with the respective metrics \(\eta_{ab} = \mathrm{diag}(-1, 1, 1, 1)\) and \(g_{\mu\nu}\).

\section{Newman-Penrose formalism}\label{sec:newpen}
We begin with a brief review of the Newman-Penrose formalism, where we use the notation and conventions of~\cite{Will:1993ns}. An important ingredient of this formalism is the assumption that a gravitational wave detector can be modeled as an ensemble of test particles, which follow the geodesics of a metric \(g_{\mu\nu}\), and that the observable effect of the wave is described by measuring the proper (geodesic) distance between different test particles. Under these assumptions the observed distances evolve according to the geodesic deviation equation
\begin{equation}\label{eqn:geodev}
u^{\rho}\lc{\nabla}_{\rho}\left(u^{\sigma}\lc{\nabla}_{\sigma}s^{\mu}\right) = -\lc{R}^{\mu}{}_{\rho\nu\sigma}u^{\rho}u^{\sigma}s^{\nu}\,,
\end{equation}
where \(u^{\mu}\) is a vector field describing the tangent vectors to a congruence of geodesics, while \(s^{\mu}\) describes the distance between nearby geodesics. Here we introduced the notation \(\lc{\nabla}\) for the covariant derivative of the Levi-Civita connection of the metric \(g_{\mu\nu}\) and \(\lc{R}^{\mu}{}_{\rho\nu\sigma}\) for its Riemann tensor. The crucial assumption that justifies the use of the Levi-Civita connection here is that the matter constituting the test particles couples only to the metric and is not influenced by other fields present in the theory.

The gravitational wave is modeled as a plane null wave describing a linear perturbation
\begin{equation}\label{eqn:mzwave}
g_{\mu\nu} = \eta_{\mu\nu} + \epsilon h_{\mu\nu}\,, \quad
h_{\mu\nu} = H_{\mu\nu}e^{i\omega u}\,,
\end{equation}
with perturbation parameter \(\epsilon \ll 1\), amplitude \(H_{\mu\nu}\) and frequency \(\omega\) propagating in the null direction \(u = x^0 - x^3\). One then introduces the complex double null basis
\begin{equation}\label{eqn:npbasis}
l = \partial_0 + \partial_3\,, \quad
n = \frac{1}{2}(\partial_0 - \partial_3)\,, \quad
m = \frac{1}{\sqrt{2}}(\partial_1 + i\partial_2)\,, \quad
\bar{m} = \frac{1}{\sqrt{2}}(\partial_1 - i\partial_2)\,.
\end{equation}
Note that not all components of the Riemann tensor are relevant for the geodesic deviation~\eqref{eqn:geodev} of an ensemble of point particles, but only the six ``electric'' components \(R_{\mu\rho\nu\sigma}u^{\rho}u^{\sigma}\) obtained by a projection onto the timelike vector \(u^{\mu}\). An equivalent description of these components is given by replacing the timelike vector \(u^{\mu}\) with the null vector \(n^{\mu}\). In the basis~\eqref{eqn:npbasis}, these components read
\begin{gather}
\Psi_2 = -\frac{1}{6}\lc{R}_{nlnl} = \frac{1}{12}\ddot{h}_{ll}\,, \quad
\Psi_3 = -\frac{1}{2}\lc{R}_{nln\bar{m}} = -\frac{1}{2}\overline{\lc{R}_{nlnm}} = \frac{1}{4}\ddot{h}_{l\bar{m}} = \frac{1}{4}\overline{\ddot{h}_{lm}}\,,\label{eqn:riemcomp}\\
\Psi_4 = -\lc{R}_{n\bar{m}n\bar{m}} = -\overline{\lc{R}_{nmnm}} = \frac{1}{2}\ddot{h}_{\bar{m}\bar{m}} = \frac{1}{2}\overline{\ddot{h}_{mm}}\,,
\quad \Phi_{22} = -\lc{R}_{nmn\bar{m}} = \frac{1}{2}\ddot{h}_{m\bar{m}}\nonumber
\end{gather}
for the wave we consider here.

We finally mention that under a change of coordinates which leaves the direction and frequency of the wave invariant, these components form a representation of the Euclidean group. This fact has been used to derive a classification scheme for gravity theories in terms of such representations, which denote the polarizations that are present in a particular theory~\cite{Eardley:1973br,Eardley:1974nw}, and which we employ in the following two sections to different classes of teleparallel gravity theories.

\section{Torsion teleparallel gravity}\label{sec:torsion}
We now apply the Newman-Penrose formalism to a class of teleparallel gravity theories developed in~\cite{Bahamonde:2017wwk}, which is an extension of ``new general relativity''~\cite{Hayashi:1979qx}, and in which torsion describes the gravitational interaction. We briefly review the teleparallel geometry in section~\ref{ssec:tfields}. The dynamics are defined in section~\ref{ssec:taction}. We derive the linearized field equations in section~\ref{ssec:tpert}. Finally, in section~\ref{ssec:twavpol} we determine the allowed polarizations of gravitational waves. A more thorough discussion of gravitational waves in the linearized theory, including a discussion of their propagation velocity, is given in~\cite{HKPU:2018mtp}.

\subsection{Dynamical fields and geometry}\label{ssec:tfields}
The fundamental fields which define the teleparallel geometry are a tetrad \(\theta^a{}_{\mu}\) and a flat Lorentz spin connection \(\omega^a{}_{b\mu}\). A number of quantities are derived from these fields. We denote by \(e_a{}^{\mu}\) the inverse tetrad, which satisfies \(\theta^a{}_{\mu}e_a{}^{\nu} = \delta_{\mu}^{\nu}\) and \(\theta^a{}_{\mu}e_b{}^{\mu} = \delta^a_b\). One further defines a metric
\begin{equation}
g_{\mu\nu} = \eta_{ab}\theta^a{}_{\mu}\theta^b{}_{\nu}\,,
\end{equation}
as well as an affine connection with coefficients
\begin{equation}
\Gamma^{\rho}{}_{\mu\nu} = e_a{}^{\rho}\left(\partial_{\nu}\theta^a_{\mu} + \omega^a{}_{b\nu}\theta^b{}_{\mu}\right)\,.
\end{equation}
This connection is metric-compatible and flat,
\begin{equation}\label{eqn:metflat}
\nabla_{\rho}g_{\mu\nu} = 0\,, \quad
\partial_{\mu}\Gamma^{\rho}{}_{\sigma\nu} - \partial_{\nu}\Gamma^{\rho}{}_{\sigma\mu} + \Gamma^{\rho}{}_{\tau\mu}\Gamma^{\tau}{}_{\sigma\nu} - \Gamma^{\rho}{}_{\tau\nu}\Gamma^{\tau}{}_{\sigma\mu} = 0\,,
\end{equation}
but in general has non-vanishing torsion
\begin{equation}
T^{\rho}{}_{\mu\nu} = \Gamma^{\rho}{}_{\nu\mu} - \Gamma^{\rho}{}_{\mu\nu} = e_a{}^{\rho}\left(\partial_{\mu}\theta^a{}_{\nu} - \partial_{\nu}\theta^a_{\mu} + \omega^a{}_{b\mu}\theta^b{}_{\nu} - \omega^a{}_{b\nu}\theta^b{}_{\mu}\right)\,.
\end{equation}
The condition~\eqref{eqn:metflat} implies that the spin connection must be of the form \(\omega^a{}_{b\mu} = \Lambda^a{}_c\partial_{\mu}\Lambda_b{}^c\) in order to have vanishing curvature, where \(\eta_{ab}\Lambda^a{}_c\Lambda^b{}_d = \eta_{cd}\) to ensure metric compatibility. We will henceforth assume these two properties throughout this section.

\subsection{Action and field equations}\label{ssec:taction}
We now come to the dynamical equations for the aforementioned fundamental field variables. These are derived from an action of the form
\begin{equation}\label{eqn:taction}
S[\theta, \omega, \chi] = S_g[\theta, \omega] + S_m[\theta, \chi]\,,
\end{equation}
where we have introduced a set \(\chi\) of matter fields. We write the variation of the matter action \(S_m\) with respect to the tetrad in the form
\begin{equation}
\delta_{\theta}S_m = -\int_M\Theta_a{}^{\mu}\delta\theta^a{}_{\mu}\,\theta\,d^4x\,,
\end{equation}
where \(\theta\) denotes the determinant of the tetrad and we introduced the energy-momentum tensor \(\Theta_a{}^{\mu}\). We assume that the matter action is invariant under local Lorentz transformations \(\theta^a{}_{\mu} \mapsto \Lambda^a{}_b\theta^b{}_{\mu}\) of the tetrad, which then implies that the energy-momentum tensor is symmetric, \(\Theta_{[\mu\nu]} = 0\).

For the gravitational part \(S_g\) of the action we choose the form
\begin{equation}
S_g[\theta, \omega] = \frac{1}{2\kappa^2}\int_M\mathcal{F}(\mathcal{T}_1, \mathcal{T}_2, \mathcal{T}_3)\theta\,d^4x\,,
\end{equation}
where the free function \(\mathcal{F}\) depends on the three scalar quantities
\begin{equation}
\mathcal{T}_1 = T^{\mu\nu\rho}T_{\mu\nu\rho}\,, \quad
\mathcal{T}_2 = T^{\mu\nu\rho}T_{\rho\nu\mu}\,, \quad
\mathcal{T}_3 = T^{\mu}{}_{\mu\rho}T_{\nu}{}^{\nu\rho}
\end{equation}
which are quadratic in the torsion.

By variation of the total action~\eqref{eqn:taction} with respect to the tetrad we obtain the gravitational field equations
\begin{equation}\label{eqn:tfeqtetrad}
\begin{split}
\kappa^2\Theta_{\mu\nu} &= \frac{1}{2}\mathcal{F}g_{\mu\nu} + 2\lc{\nabla}^{\rho}\left(\mathcal{F}_{,1}T_{\nu\mu\rho} + \mathcal{F}_{,2}T_{[\rho\mu]\nu} + \mathcal{F}_{,3}T^{\sigma}{}_{\sigma[\rho}g_{\mu]\nu}\right)\\
&\phantom{=}+ \mathcal{F}_{,1}T^{\rho\sigma}{}_{\mu}\left(T_{\nu\rho\sigma} - 2T_{[\rho\sigma]\nu}\right) - \frac{1}{2}\mathcal{F}_{,3}T^{\sigma}{}_{\sigma\rho}\left(T^{\rho}{}_{\mu\nu} + 2T_{(\mu\nu)}{}^{\rho}\right)\\
&\phantom{=}+ \frac{1}{2}\mathcal{F}_{,2}\left[T_{\mu}{}^{\rho\sigma}\left(2T_{\rho\sigma\nu} - T_{\nu\rho\sigma}\right) + T^{\rho\sigma}{}_{\mu}\left(2T_{[\rho\sigma]\nu} - T_{\nu\rho\sigma}\right)\right]\,,
\end{split}
\end{equation}
where we introduced the notation
\begin{equation}
\mathcal{F}_{,i} = \frac{\partial\mathcal{F}}{\partial\mathcal{T}_i}\,, \quad i = 1, 2, 3\,,
\end{equation}
and already used the fact that \(\Theta_{\mu\nu}\) is symmetric. The antisymmetric part of these field equations is identical to the field equations obtained by variation with respect to the spin connection. Note that we must restrict ourselves to variations of the form
\begin{equation}
\delta\omega^a{}_{b\mu} = \partial_{\mu}\upsilon^a{}_b + \omega^a{}_{c\mu}\upsilon^c{}_b - \omega^c{}_{b\mu}\upsilon^a{}_c
\end{equation}
with \(\upsilon^{(ab)} = 0\), in order to preserve that \(\omega\) is a flat Lorentz spin connection~\cite{Golovnev:2017dox}.

\subsection{Perturbation ansatz and linearized field equations}\label{ssec:tpert}
The gravitational wave we describe here will be a linear perturbation around a fixed vacuum solution, and so we will assume \(\Theta_{\mu\nu} = 0\). Assuming that the background solution is given by a Minkowski metric, expressed by a diagonal tetrad and vanishing spin connection, we can write this perturbation as
\begin{equation}
\theta^a{}_{\mu} = \Delta^a{}_{\mu} + \epsilon\tau^a{}_{\mu}\,, \quad
e_a{}^{\mu} = (\Delta^{-1})_a{}^{\mu} - \epsilon\tau_a{}^{\mu}\,, \quad
\omega^a{}_{b\mu} = \epsilon\partial_{\mu}\lambda^a{}_b\,,
\end{equation}
where \(\epsilon\) is the perturbation parameter and \(\lambda^{(ab)} = 0\). Here we used the notation \(\Delta^a{}_{\mu}\) for the diagonal tetrad, which in Cartesian coordinates has the components \(\Delta^a{}_{\mu} = \mathrm{diag}(1, 1, 1, 1) = \delta^a{}_{\mu}\). Using this ansatz the torsion tensor reads
\begin{equation}
T^{\rho}{}_{\mu\nu} = 2\epsilon(\Delta^{-1})_a{}^{\rho}\left(\partial_{[\mu}\tau^a{}_{\nu]} - \Delta^b{}_{[\nu}\partial_{\mu]}\lambda^a{}_b\right) + \mathcal{O}(\epsilon^2)\,.
\end{equation}
Employing the convention that Latin and Greek indices can be exchanged by making use of the tetrad, we can also write the torsion in the form
\begin{equation}
T^{\rho}{}_{\mu\nu} = 2\epsilon\left(\partial_{[\mu}\tau^{\rho}{}_{\nu]} - \partial_{[\mu}\lambda^{\rho}{}_{\nu]}\right) + \mathcal{O}(\epsilon^2)\,.
\end{equation}
We see that the torsion is already of linear order in the perturbations. The scalar quantities \(\mathcal{T}_i\) are thus of quadratic order \(\mathcal{O}(\epsilon^2)\). From a Taylor expansion of the Lagrangian function \(\mathcal{F}\) then follows that
\begin{equation}
\mathcal{F} = \mathcal{F}|_{\mathcal{T}_i = 0} + \mathcal{O}(\epsilon^2) = F + \mathcal{O}(\epsilon^2)\,, \quad
\mathcal{F}_{,i} = \mathcal{F}_{,i}|_{\mathcal{T}_i = 0} + \mathcal{O}(\epsilon^2) = F_{,i} + \mathcal{O}(\epsilon^2)
\end{equation}
so that only the constant background value of \(\mathcal{F}\) and its derivative contribute to the linearized field equations. We denote these constant background values by \(F\) and \(F_{,i}\), respectively. Finally, up to the required perturbation orders the metric and its Levi-Civita connection can be written as
\begin{equation}
g_{\mu\nu} = \eta_{\mu\nu} + 2\epsilon\tau_{(\mu\nu)} + \mathcal{O}(\epsilon^2)\,, \quad
\lc{\nabla}_{\mu} = \partial_{\mu} + \mathcal{O}(\epsilon)\,.
\end{equation}
Inserting these expansions into the field equations~\eqref{eqn:tfeqtetrad} we find that at the zeroth order they reduce to \(F\eta_{\mu\nu} = 0\). For consistency we therefore assume \(F = 0\), i.e., we neglect the presence of a cosmological constant. Introducing the notation \(s_{\mu\nu} = \tau_{(\mu\nu)}\) and \(a_{\mu\nu} = \tau_{[\mu\nu]} - \lambda_{\mu\nu}\), as well as \(E_{\mu\nu}\) for the gravitational side of the field equations, we obtain the linearized vacuum field equations
\begin{equation}\label{eqn:tlinvaceom}
\begin{split}
0 = E_{\mu\nu} &= \partial^{\rho}\left[2(2F_{,1} - F_{,2})\partial_{[\nu}a_{\rho]\mu} + (2F_{,2} + F_{,3})\partial_{\mu}a_{\rho\nu}\right]\\
&\phantom{=}+ 2\partial^{\rho}\left[(2F_{,1} + F_{,2})\partial_{[\rho}s_{\nu]\mu} + F_{,3}\left(\eta_{\mu[\nu}\partial_{\rho]}s^{\sigma}{}_{\sigma} - \partial^{\sigma}s_{\sigma[\rho}\eta_{\nu]\mu}\right)\right]\,,
\end{split}
\end{equation}
These are the equations we will use.

\subsection{Polarization of gravitational waves}\label{ssec:twavpol}
\begin{figure}[htb]
\centerline{\includegraphics[width=0.8\textwidth]{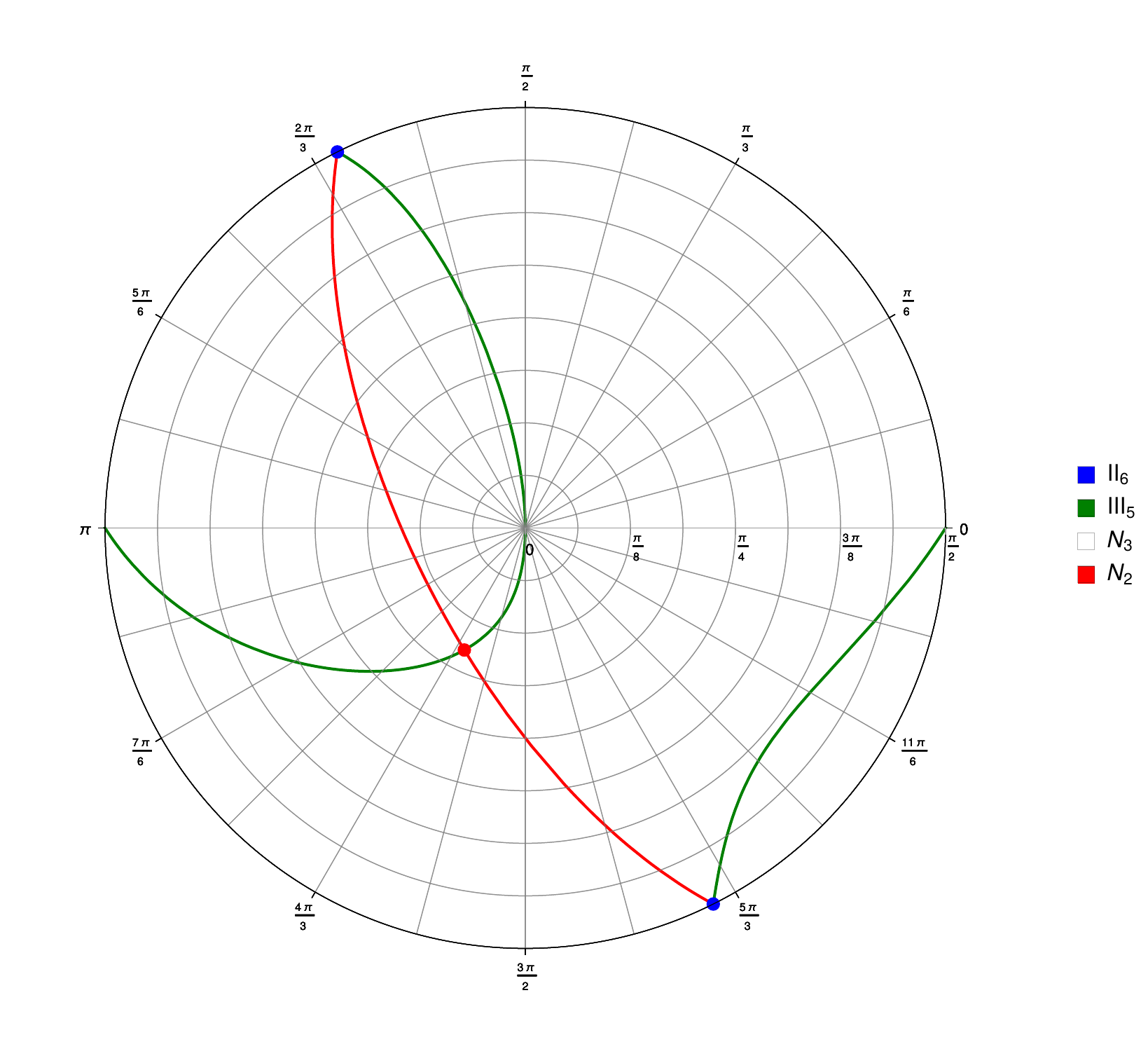}}
\caption{Visualization of the parameter space.}
\label{fig:ngrpol}
\end{figure}

We now consider a plane wave of the form
\begin{equation}\label{eqn:tzwave}
s_{\mu\nu} = S_{\mu\nu}e^{i\omega u}\,, \quad
a_{\mu\nu} = A_{\mu\nu}e^{i\omega u}\,,
\end{equation}
with amplitudes \(S_{\mu\nu}\) and \(A_{\mu\nu}\) for a single Fourier mode of frequency \(\omega\). Here \(u = x^0 - x^3\) is the retarded time. In the Newman-Penrose basis~\eqref{eqn:npbasis} the relevant components of the linearized field equations~\eqref{eqn:tlinvaceom} applied to this wave are
\begin{subequations}\label{eqn:tnp_wave}
\begin{align}
0 &= E_{nn} = (2F_{,1} + F_{,2} + F_{,3})\ddot{s}_{nl} + 2F_{,3}\ddot{s}_{m\bar{m}} + (2F_{,1} + F_{,2} + F_{,3})\ddot{a}_{nl}\,,\label{eqn:tnp_nn}\\
0 &= E_{mn} = \overline{E_{\bar{m}n}} = (2F_{,1} + F_{,2})\ddot{s}_{ml} + (2F_{,1} - F_{,2})\ddot{a}_{ml}\,,\label{eqn:tnp_mn}\\
0 &= E_{nm} = \overline{E_{n\bar{m}}} = -F_{,3}\ddot{s}_{ml} + (2F_{,2} + F_{,3})\ddot{a}_{ml}\,,\label{eqn:tnp_nm}\\
0 &= E_{m\bar{m}} = E_{\bar{m}m}= -F_{,3}\ddot{s}_{ll}\,,\label{eqn:tnp_mbm}\\
0 &= E_{ln} = (2F_{,1} + F_{,2})\ddot{s}_{ll}\,.\label{eqn:tnp_ln}
\end{align}
\end{subequations}
Recall that the metric perturbations are given by \(h_{\mu\nu} = 2s_{\mu\nu}\). For the electric components~\eqref{eqn:riemcomp} of the Riemann tensor we thus find the following possibilities:
\begin{itemize}
\item[\tikz{\filldraw[fill=blue,draw=black](-0.15,-0.15) rectangle (0.15,0.15);}]
\(2F_{,1} + F_{,2} = F_{,3} = 0\):
None of the modes~\eqref{eqn:riemcomp} is restricted by the linearized field equations. Theories satisfying these conditions belong to the \(\mathrm{E}(2)\) class \(\mathrm{II}_6\), shown in blue in figure~\ref{fig:ngrpol}.

\item[\tikz{\filldraw[fill=green!50!black,draw=black](-0.15,-0.15) rectangle (0.15,0.15);}]
\(2F_{,1}(F_{,2} + F_{,3}) + F_{,2}^2 = 0\) and \(2F_{,1} + F_{,2} + F_{,3} \neq 0\):
In this case the field equations enforce \(\Psi_2 = 0\), so that there is no longitudinal mode. All other modes are unrestricted. Theories of this type belong to the \(\mathrm{E}(2)\) class \(\mathrm{III}_5\). This case is represented by the green line in figure~\ref{fig:ngrpol}.

\item[\tikz{\filldraw[fill=white,draw=black](-0.15,-0.15) rectangle (0.15,0.15);}]
\(2F_{,1}(F_{,2} + F_{,3}) + F_{,2}^2 \neq 0\) and \(2F_{,1} + F_{,2} + F_{,3} \neq 0\):
From the field equations follows \(\Psi_2 = \Psi_3 = 0\), while the breathing mode \(\Phi_{22}\) and tensor modes \(\Psi_4\) are unrestricted. This wave has the \(\mathrm{E}(2)\) class \(\mathrm{N}_3\). Almost all points of the parameter space, shown in white in figure~\ref{fig:ngrpol}, belong to this class.

\item[\tikz{\filldraw[fill=red,draw=black](-0.15,-0.15) rectangle (0.15,0.15);}]
\(2F_{,1} + F_{,2} + F_{,3} = 0\) and \(F_{,3} \neq 0\):
The only mode which is allowed to be nonzero is given by the two tensor polarizations \(\Psi_4\). The \(\mathrm{E}(2)\) class of this wave is \(\mathrm{N}_2\). This case is shown as a red line in figure~\ref{fig:ngrpol}. Note in particular that TEGR, marked as a red point, belongs to this class.
\end{itemize}

The different \(\mathrm{E}(2)\) classes are summarized in figure~\ref{fig:ngrpol}, where the radial coordinate \(\vartheta\) and angular coordinate \(\varphi\) are defined such that
\begin{equation}
F_{,1} = C\sin\vartheta\cos\varphi\,, \quad
F_{,2} = C\sin\vartheta\sin\varphi\,, \quad
F_{,3} = C\cos\vartheta\,,
\end{equation}
where \(C = \sqrt{F_{,1}^2 + F_{,2}^2 + F_{,3}^2}\) is assumed to be nonzero.

\section{Non-metricity teleparallel gravity}\label{sec:nonmet}
We now turn our focus to a class of symmetric teleparallel gravity theories extending the model shown in~\cite{BeltranJimenez:2017tkd,BeltranJimenez:2018vdo}, in which gravity is mediated by the non-metricity. We discuss the dynamical fields in section~\ref{ssec:nfields}, action and field equations in section~\ref{ssec:naction}, linearized field equations in section~\ref{ssec:npert}, and finally the gravitational wave polarizations in section~\ref{ssec:nwavpol}. A full discussion of the speed and polarization of gravitational waves in the linearized theory is given in~\cite{HPSU:2018stp}.

\subsection{Dynamical fields and geometry}\label{ssec:nfields}
We start our discussion of symmetric teleparallel gravity with an overview of the dynamical fields. These are given by a Lorentzian metric \(g_{\mu\nu}\) and an affine connection \(\Gamma^{\rho}{}_{\mu\nu}\). The connection is assumed to be flat and symmetric, i.e., it has vanishing torsion and curvature,
\begin{equation}\label{eqn:symflat}
T^{\rho}{}_{\mu\nu} = 2\Gamma^{\rho}{}_{[\nu\mu]} = 0\,, \quad
\partial_{\mu}\Gamma^{\rho}{}_{\sigma\nu} - \partial_{\nu}\Gamma^{\rho}{}_{\sigma\mu} + \Gamma^{\rho}{}_{\tau\mu}\Gamma^{\tau}{}_{\sigma\nu} - \Gamma^{\rho}{}_{\tau\nu}\Gamma^{\tau}{}_{\sigma\mu} = 0\,,
\end{equation}
but in general non-vanishing non-metricity
\begin{equation}
Q_{\rho\mu\nu} = \nabla_{\rho}g_{\mu\nu}\,.
\end{equation}
It follows from the condition~\eqref{eqn:symflat} that the affine connection must be of the form \(\Gamma^{\rho}{}_{\mu\nu} = (\Lambda^{-1})^{\rho}{}_{\sigma}\partial_{\nu}\Lambda^{\sigma}{}_{\mu}\) in order to have vanishing curvature, where \(\partial_{[\nu}\Lambda^{\sigma}{}_{\mu]} = 0\) to ensure the symmetry of the connection. We will henceforth assume these two properties throughout this section.

\subsection{Action and field equations}\label{ssec:naction}
We now derive the dynamics for the fields detailed above. The action we consider has a similar structure to the action given in section~\ref{ssec:taction} and reads
\begin{equation}\label{eqn:naction}
S[g, \Gamma, \chi] = S_g[g, \Gamma] + S_m[g, \chi]\,,
\end{equation}
where we used \(\chi\) once again to denote a set of matter fields. By variation of the matter action \(S_m\) with respect to the metric we obtain the usual energy-momentum tensor
\begin{equation}
\delta_{\theta}S_m = -\frac{1}{2}\int_M\Theta^{\mu\nu}\delta g_{\mu\nu}\sqrt{-g}\,d^4x\,,
\end{equation}
where \(g\) denotes the determinant of the metric \(g_{\mu\nu}\).

The gravitational part \(S_g\) of the action we consider here takes the form
\begin{equation}
S_g[g, \Gamma] = \int_M\mathcal{F}(\mathcal{Q}_1, \mathcal{Q}_2, \mathcal{Q}_3, \mathcal{Q}_4, \mathcal{Q}_5)\sqrt{-g}\,d^4x\,,
\end{equation}
where the Lagrangian is given by a free function \(\mathcal{F}\) of the five scalar quantities
\begin{gather}
\mathcal{Q}_1 = Q^{\mu\nu\rho}Q_{\mu\nu\rho}\,, \quad
\mathcal{Q}_2 = Q^{\mu\nu\rho}Q_{\rho\mu\nu}\,, \quad
\mathcal{Q}_3 = Q^{\rho\mu}{}_{\mu}Q_{\rho\nu}{}^{\nu}\,, \nonumber\\
\mathcal{Q}_4 = Q^{\mu}{}_{\mu\rho}Q_{\nu}{}^{\nu\rho}\,, \quad
\mathcal{Q}_5 = Q^{\mu}{}_{\mu\rho}Q^{\rho\nu}{}_{\nu}\,,
\end{gather}
which are quadratic in the non-metricity.

By variation of the total action~\eqref{eqn:naction} we obtain the field equations
\begin{equation}\label{eqn:nfeqmetric}
\begin{split}
\kappa^2\Theta_{\mu\nu} &= -2\lc{\nabla}_{\rho}\left(\mathcal{F}_{,1}Q^{\rho}{}_{\mu\nu} + \mathcal{F}_{,2}Q_{(\mu\nu)}{}^{\rho} + \mathcal{F}_{,3}Q^{\rho\sigma}{}_{\sigma}g_{\mu\nu} + \mathcal{F}_{,4}Q^{\sigma}{}_{\sigma(\mu}\delta_{\nu)}^{\rho}\right)\\
&\phantom{=}- \lc{\nabla}_{\rho}\left[\mathcal{F}_{,5}\left(Q_{\sigma}{}^{\sigma\rho}g_{\mu\nu} + \delta^{\rho}_{(\mu}Q_{\nu)\sigma}{}^{\sigma}\right)\right] + \frac{1}{2}\mathcal{F}g_{\mu\nu} - \mathcal{F}_{,3}Q_{\mu\rho}{}^{\rho}Q_{\nu\sigma}{}^{\sigma}\\
&\phantom{=}+ \mathcal{F}_{,1}\left(2Q^{\rho\sigma}{}_{\mu}Q_{\sigma\rho\nu} - Q_{\mu}{}^{\rho\sigma}Q_{\nu\rho\sigma} - 2Q^{\rho\sigma}{}_{(\mu}Q_{\nu)\rho\sigma}\right)\\
&\phantom{=}+ \mathcal{F}_{,2}\left(Q^{\rho\sigma}{}_{\mu}Q_{\rho\sigma\nu} - Q_{\mu}{}^{\rho\sigma}Q_{\nu\rho\sigma} - Q^{\rho\sigma}{}_{(\mu}Q_{\nu)\rho\sigma}\right)\\
&\phantom{=}+ \mathcal{F}_{,4}\left[Q_{\rho}{}^{\rho\sigma}\left(Q_{\sigma\mu\nu} - 2Q_{(\mu\nu)\sigma}\right) + Q^{\rho}{}_{\rho\mu}Q^{\sigma}{}_{\sigma\nu} - Q^{\rho}{}_{\rho(\mu}Q_{\nu)\sigma}{}^{\sigma}\right]\\
&\phantom{=}+ \frac{1}{2}\mathcal{F}_{,5}\left[Q^{\rho\sigma}{}_{\sigma}\left(Q_{\rho\mu\nu} - 2Q_{(\mu\nu)\rho}\right) - Q_{\mu\rho}{}^{\rho}Q_{\nu\sigma}{}^{\sigma}\right]\,,
\end{split}
\end{equation}
where we adapted our notation such as to denote
\begin{equation}
\mathcal{F}_{,i} = \frac{\partial\mathcal{F}}{\partial\mathcal{Q}_i}\,, \quad i = 1, \ldots, 5\,.
\end{equation}
The second set of field equations is obtained by variation with respect to the flat, symmetric connection, where we allow only variations of the form \(\delta\Gamma^{\rho}{}_{\mu\nu} = \nabla_{\mu}\nabla_{\nu}\upsilon^{\rho}\) in order to preserve these two conditions. The resulting field equations can be obtained from the metric field equations by taking their covariant divergence with respect to the Levi-Civita covariant derivative \(\lc{\nabla}\).

\subsection{Perturbation ansatz and linearized field equations}\label{ssec:npert}
We now restrict ourselves to vacuum solutions, and henceforth assume \(\Theta_{\mu\nu} = 0\). For the background solution we consider a Minkowski metric in the coincident gauge \(\Gamma^{\rho}{}_{\mu\nu} = 0\). The linear perturbation is thus given by
\begin{equation}
g_{\mu\nu} = \eta_{\mu\nu} + \epsilon h_{\mu\nu}\,, \quad
\Gamma^{\rho}{}_{\mu\nu} = \epsilon\partial_{\mu}\partial_{\nu}\xi^{\rho}\,,
\end{equation}
where \(\Gamma^{\rho}{}_{\mu\nu}\) is chosen such that the connection is flat and symmetric. We find that the non-metricity
\begin{equation}
Q_{\rho\mu\nu} = \epsilon\partial_{\rho}(h_{\mu\nu} - 2\partial_{(\mu}\xi_{\nu)}) + \mathcal{O}(\epsilon^2)
\end{equation}
is of linear order in the perturbations and depends only on the quantities \(b_{\mu\nu} = h_{\mu\nu} - 2\partial_{(\mu}\xi_{\nu)}\). It thus follows that the quantities \(\mathcal{Q}_i\) are of quadratic order \(\mathcal{O}(\epsilon^2)\). Performing a Taylor expansion
\begin{equation}
\mathcal{F} = \mathcal{F}|_{\mathcal{Q}_i = 0} + \mathcal{O}(\epsilon^2) = F + \mathcal{O}(\epsilon^2)\,, \quad
\mathcal{F}_{,i} = \mathcal{F}_{,i}|_{\mathcal{Q}_i = 0} + \mathcal{O}(\epsilon^2) = F_{,i} + \mathcal{O}(\epsilon^2)
\end{equation}
of the Lagrangian function \(\mathcal{F}\), we thus see that only the constant background values \(F\) and \(F_{,i}\) contribute to the linearized field equations. Inserting the perturbation ansatz into the field equations~\eqref{eqn:nfeqmetric} we obtain the zeroth order equation \(F\eta_{\mu\nu} = 0\). As in the torsion case we restrict ourselves to theories satisfying \(F = 0\), and hence neglect the presence of a cosmological constant. Denoting the gravitational side of the field equations by \(E_{\mu\nu}\), the linearized vacuum field equations can be written as
\begin{equation}\label{eqn:nlinvaceom}
\begin{split}
0 &= 2F_{,1}\square b_{\mu\nu} + (F_{,2} + F_{,4})\eta^{\alpha\sigma}\left(\partial_{\alpha}\partial_{\mu}b_{\sigma\nu} + \partial_{\alpha}\partial_{\nu}b_{\sigma\mu}\right) + 2F_{,3}\eta_{\mu\nu}\eta^{\tau\omega}\square b_{\tau\omega}\\
&\phantom{=}+ F_{,5}\eta_{\mu\nu}\eta^{\omega\gamma}\eta^{\alpha\sigma}\partial_{\alpha}\partial_{\omega}b_{\sigma\gamma} + F_{,5}\eta^{\omega\sigma}\partial_{\mu}\partial_{\nu}b_{\omega\sigma}\,.
\end{split}
\end{equation}
We can now use these equations to derive the gravitational wave solutions.

\subsection{Polarization of gravitational waves}\label{ssec:nwavpol}
\begin{figure}[htb]
\centerline{\includegraphics[width=0.8\textwidth]{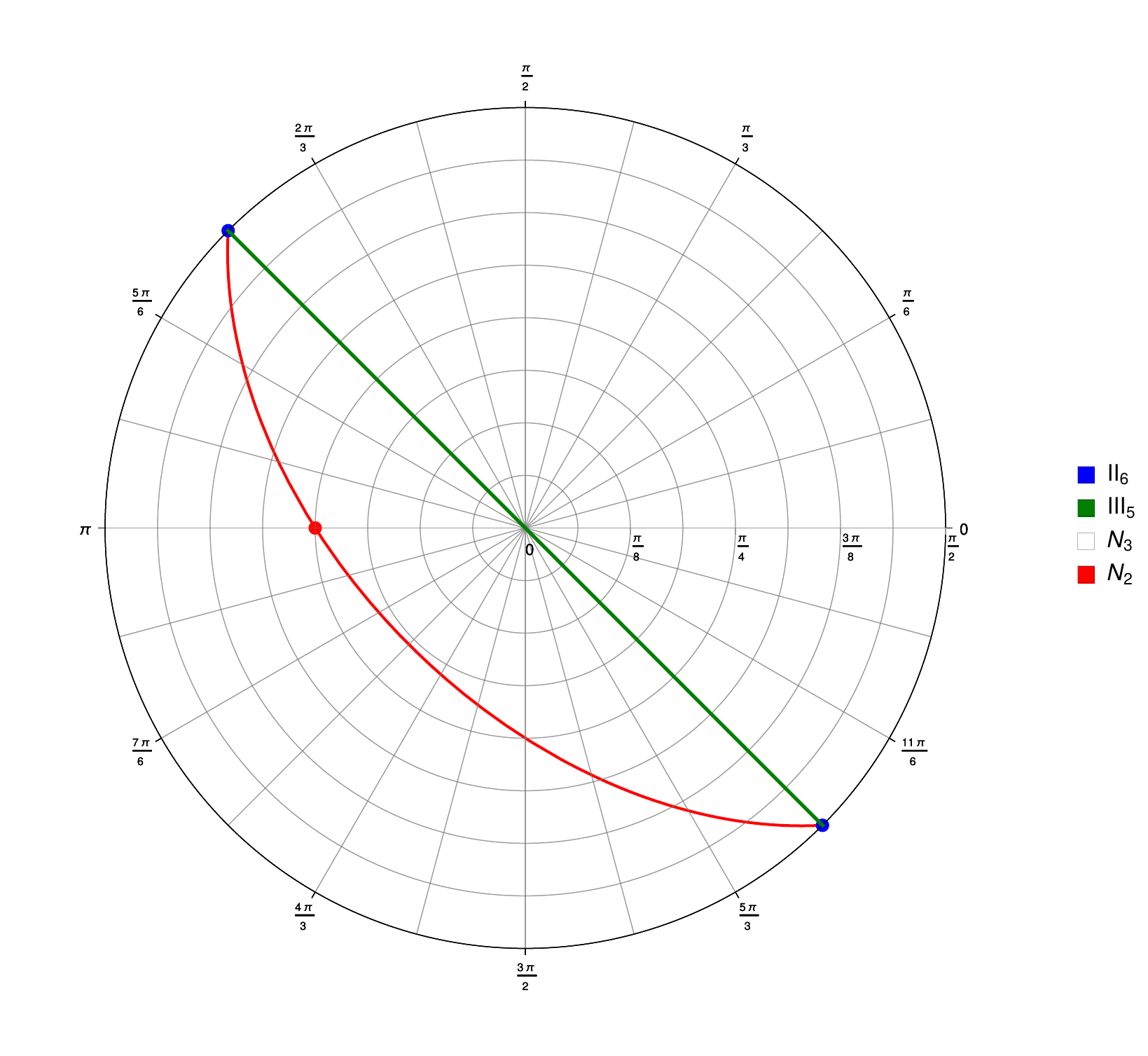}}
\caption{Visualization of the parameter space.}
\label{fig:nmpol}
\end{figure}

The gravitational wave we consider now is given by a single Fourier mode
\begin{equation}\label{eqn:nzwave}
b_{\mu\nu} = B_{\mu\nu}e^{i\omega u}
\end{equation}
with amplitude \(B_{\mu\nu}\) and frequency \(\omega\), where \(u = x^0 - x^3\) again denotes the retarded time. Inserting this wave into the linearized field equations~\eqref{eqn:nlinvaceom} and expressing the components of \(E_{\mu\nu}\) in the Newman-Penrose basis~\eqref{sec:newpen} yield the only non-vanishing and linearly independent equations
\begin{subequations}\label{eqn:nnp_wave}
\begin{align}
0 &= E_{nn} = 2F_{,5}\ddot{b}_{m\bar{m}} - 2(F_{,2} + F_{,4} + F_{,5})\ddot{b}_{ln}\,,\label{eqn:nnp_nn}\\
0 &= E_{nm} = \overline{E_{n\bar{m}}} = -(F_{,2} + F_{,4})\ddot{b}_{lm}\,,\label{eqn:nnp_nm}\\
0 &= E_{m\bar{m}} = F_{,5}\ddot{b}_{ll}\,,\label{eqn:nnp_mbm}\\
0 &= E_{ln} = -(F_{,2} + F_{,4})\ddot{b}_{ll}\,.\label{eqn:nnp_ln}
\end{align}
\end{subequations}
Depending on the coefficients \(F_{,i}\) we then find the following possibilities for the allowed electric Riemann tensor components~\eqref{eqn:riemcomp}:
\begin{itemize}
\item[\tikz{\filldraw[fill=blue,draw=black](-0.15,-0.15) rectangle (0.15,0.15);}]
\(F_{,2} + F_{,4} = F_{,5} = 0\):
In this case the linearized field equations are satisfied identically for an arbitrary plane null wave and there are no restrictions on the allowed polarizations. Theories of this type belong to the \(\mathrm{E}(2)\) class \(\mathrm{II}_6\), shown in blue in figure~\ref{fig:nmpol}.

\item[\tikz{\filldraw[fill=green!50!black,draw=black](-0.15,-0.15) rectangle (0.15,0.15);}]
\(F_{,2} + F_{,4} = 0\) and \(F_{,5} \neq 0\):
The longitudinal mode \(\Psi_2\) is prohibited in this case, while the remaining modes are unrestricted. This corresponds to the \(\mathrm{E}(2)\) class \(\mathrm{III}_5\), shown as a green line in figure~\ref{fig:nmpol}.

\item[\tikz{\filldraw[fill=white,draw=black](-0.15,-0.15) rectangle (0.15,0.15);}]
\(F_{,2} + F_{,4} \neq 0\) and \(F_{,2} + F_{,4} + F_{,5} \neq 0\):
The only allowed modes in this case are the breathing mode \(\Phi_{22}\) and the two tensor modes \(\Psi_4\), while the longitudinal mode \(\Psi_2\) and the two vector modes \(\Psi_3\) are prohibited. These theories have the \(\mathrm{E}(2)\) class \(\mathrm{N}_3\), occupying most of the parameter space shown in figure~\ref{fig:nmpol} in white.

\item[\tikz{\filldraw[fill=red,draw=black](-0.15,-0.15) rectangle (0.15,0.15);}]
\(F_{,2} + F_{,4} + F_{,5} = 0\) and \(F_{,5} \neq 0\):
The only allowed polarizations are the two tensor modes \(\Psi_2\). These theories belong to the \(\mathrm{E}(2)\) class \(\mathrm{N}_2\), shown as a red line in figure~\ref{fig:nmpol}. This includes STEGR, marked as a red point.
\end{itemize}

We have summarized all possible \(\mathrm{E}(2)\) classes in figure~\ref{fig:nmpol}. The radial coordinate \(\vartheta\) and angular coordinate \(\varphi\) are defined via
\begin{equation}
F_{,2} = C\sin\vartheta\cos\varphi\,, \quad
F_{,4} = C\sin\vartheta\sin\varphi\,, \quad
F_{,5} = C\cos\vartheta\,,
\end{equation}
where the normalization factor \(C = \sqrt{F_{,2}^2 + F_{,4}^2 + F_{,5}^2}\) is assumed to be nonzero. In the case \(C = 0\) the linearized field equations do not pose any restrictions on the gravitational wave polarizations, so that the \(\mathrm{E}(2)\) class is \(\mathrm{II}_6\).

\section{Conclusion}\label{sec:conclusion}
We have derived the allowed polarizations of gravitational waves for two general classes of teleparallel gravity theories, which are based on torsion and on non-metricity. Our results show that in both cases the theories can be divided into four classes, depending on three coefficients in the Taylor expansion of their Lagrangians, corresponding to the \(\mathrm{E}(2)\) classes \(\mathrm{N}_2\), \(\mathrm{N}_3\), \(\mathrm{III}_5\) and \(\mathrm{II}_6\), where the subscript denotes the number of gravitational wave polarizations.

We remark that since we only considered the linearized field equations, any higher order terms in the Lagrangian do not contribute to our result. Theories with higher order corrections terms are therefore indistinguishable from the ones presented here. Another interesting aspect is the presence of regions in the parameter spaces of the investigated classes of theories in which the \(E(2)\) class becomes \(\mathrm{N}_2\) as in general relativity. Theories of this type are indistinguishable from general relativity by observing their gravitational wave polarizations.

One possible extension of this work is to consider additional fields, such as one or multiple scalar fields, non-minimally coupled to torsion~\cite{Hohmann:2018rwf,Hohmann:2018vle,Hohmann:2018dqh,Hohmann:2018ijr} or to non-metricity~\cite{Jarv:2018bgs,Runkla:2018xrv}. Another potential direction for future research is to consider the propagation of gravitational waves on more general background geometries, such as a Friedmann-Lema\^{i}tre-Robertson-Walker metric, which is relevant for describing gravitational waves in expanding cosmologies.

\begin{acknowledgments}
The author thanks Martin Kr\v{s}\v{s}\'ak, Christian Pfeifer, Jackson Levi Said and Ulbossyn Ualikhanova for helpful discussions. He acknowledges the full financial support of the Estonian Ministry for Education and Science through the Institutional Research Support Project IUT02-27 and Startup Research Grant PUT790, as well as the European Regional Development Fund through the Center of Excellence TK133 ``The Dark Side of the Universe''. Participation in the conference ``The Third Zeldovich Meeting'' was supported by CANTATA, supported by COST (European Cooperation in Science and Technology), through the ITC conference grant COST-ITCCG-CA15117-398.
\end{acknowledgments}

\bibliography{zwaves}
\end{document}